# TMA diffusion process in PMMA thin films during sequential infiltration synthesis: *in situ* dynamic spectroscopic ellipsometry

Elena Cianci*,#, Daniele Nazzari#,§, Gabriele Seguini#, and Michele Perego#

#IMM-CNR, Unit of Agrate Brianza, Via C. Olivetti 2, 20864 Agrate Brianza (MB), Italy
§Department of Physics, Università Statale di Milano, Via G. Celoria 16, 20133, Milano (Mi), Italy

*KEYWORDS: In situ dynamic spectroscopic ellipsometry, Sequential Infiltration Synthesis (SIS), PMMA, PS, TMA diffusion, $Al_2O_3$.*

**ABSTRACT:** Sequential infiltration synthesis (SIS) provides a successful route to grow inorganic materials into polymeric films by penetrating of gaseous precursors into the polymer, both in order to enhance the functional properties of the polymer creating an organic-inorganic hybrid material, and to fabricate inorganic nanostructures when infiltrating in patterned polymer films or in self-assembled block copolymers. A SIS process consists in a controlled sequence of metal organic precursor and co-reactant vapor exposure cycles of the polymer films in an atomic layer deposition (ALD) reactor. In this work, we present a study of the SIS process of alumina using trimethylaluminum (TMA) and $H_2O$ in various polymer films using *in situ* dynamic spectroscopic ellipsometry (SE). *In situ* dynamic SE enables time-resolved monitoring of the swelling of the polymer, which is relevant to the diffusion and retain of the metal precursor into the polymer itself. Different swelling behaviour of poly(methylmethacrylate) (PMMA) and polystyrene (PS) was observed when exposed to TMA vapor. PMMA films swell more significantly than PS films do, resulting in very different infiltrated $Al_2O_3$ thickness after polymer removal in $O_2$ plasma. PMMA films reach different swollen states upon TMA exposure and reaction with $H_2O$, depending on the TMA dose and on the purge duration after TMA exposure, which correspond to different amounts of metal precursor retained inside the polymer and converted to alumina. Diffusion coefficients of TMA in PMMA were extracted investigating the swelling of pristine PMMA films during TMA infiltration and shown to be dependent on polymer molecular weight. *In situ* dynamic SE monitoring allows to control the SIS process tuning it from an ALD-like process for long purge to a chemical vapour deposition - like process selectively confined inside the polymer films.

## I. INTRODUCTION

Sequential infiltration synthesis (SIS) has been demonstrated to be a viable preparation route for hybrid inorganic-organic materials, by penetrating of gaseous metal precursors into polymer films, fibers, foams, or biomaterials[1–5]. The incorporation of inorganic materials into organic matrices offers the possibility to enhance the mechanical properties of the final composite[6–8], to improve its chemical etch resistance[9,10], to tune its optical properties[11], or its oil sorption capability for efficient extraction of oil from water[3]. In particular, infiltration in lithographically patterned polymeric thin films[12,13] or in self-assembled block copolymer (BCP) films[9,14,15] allows the formation of inorganic nanostructures, once the polymer matrix is removed by an $O_2$ plasma or annealing step. Simply replicating the patterned polymer or BCP templates, metal oxide ($Al_2O_3$, $TiO_2$, $SiO_2$, or $ZnO$[14]) and metal (W)[15] nano-architectures have been directly fabricated, for applications as sub-20 nm enhanced lithographic masks or as active elements in advanced nanoscale electronic devices.

Typical SIS process consists in a controlled sequence of metal organic precursor and co-reactant vapor exposures of the organic samples in an atomic layer deposition (ALD) reactor, intercalating appropriate purging cycles of inert gas to remove unreacted molecules or reaction by-products. The process is quite complex involving chemical and physical reactions of the precursors among themselves and with the host matrix. In particular, diffusion of precursors throughout the infiltrated polymer films plays a fundamental role in the kinetic of the infiltration process and need to be monitored in real time to get a comprehensive picture of the specific SIS process under investigation. In order to achieve optimal control of the infiltration process, *in situ* analysis techniques have been tailored to characterize how the polymer film is modified as a consequence of the infiltration process itself, achieving information on the evolution of the sample during the process.

The most studied SIS process has been the infiltration of trimethylaluminum (TMA) in combination with $H_2O$ for the synthesis of $Al_2O_3$ into polymeric films, and *in situ* quartz crystal microbalance (QCM) measurements and *in situ* Fourier Transform Infrared (FTIR) Spectroscopy[12,16–20] have been the mostly used *in situ* techniques for investigating the infiltration mechanism. Both characterization techniques demonstrated the dependence of effective infiltration on the presence of intermolecular interactions between the metalorganic precursor and some functional groups in the polymer, and the influence of process parameters as temperature, precursor exposure, purge duration, and pulse sequence. *In situ* QCM showed different alumina mass uptake per cycle in different polymers and allowed displaying mass uptake and loss during TMA dosing and purging respectively. In the case of polymethylmethacrylate (PMMA) films, FTIR analysis evidenced that the interaction between TMA and C=O and C-O-R moieties enables to form a reversible adduct C=O···Al(CH$_3$)$_3$ for the subsequent nucleation of $Al_2O_3$[17–19]. For a PS-*b*-PMMA BCPs made of PMMA and polystyrene (PS), TMA infiltration process discriminates PMMA respect to PS domains where no TMA-phyilic groups are available, allowing to selectively grow alumina only in PMMA volume and to create inorganic nanostructures as replicas of the BCP template [14, 20–22]. Anyway, a clear and comprehensive picture of the different phenomena occurring during a SIS process is still missing.

In this work, we investigated the infiltration process of TMA and $H_2O$ in PMMA and PS films, as building blocks of PS-b-PMMA BCPs, using *in situ* dynamic spectroscopic ellipsometry (SE). Ellipsometry is a non-invasive optical technique that has been widely used for studies of polymer layers under several conditions[24–26] and in combination with ALD processes[27–29]. *In situ* dynamic SE allows the continuous acquisition of information about the changes of the thickness and refractive index n of the polymer film during infiltration, without the need of *ad hoc* samples, as for *in situ* QCM measurements, and without interrupting the process, both over subsequent cycles and during the different steps of a single cycle, on a shorter time scale than in *in situ* FTIR spectroscopy analysis. Through *in situ* dynamic ellipsometric analysis of polymer infiltration, quantitative information about TMA diffusivity in pristine PMMA matrix were obtained, gaining further insight into the process kinetics, whose comprehension is required for process optimization and for extension of the SIS methodology to the synthesis of new materials and to other polymeric matrices.

## II. EXPERIMENTAL SECTION

PMMA ($M_n$ = 14 kg·mol$^{-1}$, PDI = 1.2 and $M_n$ = 3.9 kg·mol$^{-1}$, PDI = 1.2) and PS ($M_n$ = 13 kg·mol$^{-1}$, PDI = 1.1) thin films of different thickness (from 8 to 100 nm) were prepared by adjusting the concentration of polymer-toluene solutions. The polymeric films were deposited by spin coating on top of a (100) Si substrate, cleaned in Piranha solution ($H_2SO_4$/$H_2O_2$, 3/1 vol. ratio) at 80°C for 40 min in order to increase the hydroxyl groups surface density. Then, the samples were rinsed with 2-propanol in ultrasonic bath and $N_2$ dried. Subsequently, the films were thermally annealed at 290°C for 900 s in $N_2$ atmosphere by means of a Rapid Thermal Process (RTP) to promote the grafting of the polymer chains to the substrate surface[30–32]. After RTP, a subset of samples was sonicated in toluene, in order to remove the non-grafted polymer fraction, obtaining very thin films.

The samples were then loaded in a cross flow commercial ALD reactor (Savannah 200, Ultratech Cambridge NanoTech.), thermalized at 90°C for 30 min under 100 sccm $N_2$ flow at a 0.6 Torr chamber pressure, before starting the infiltration process. TMA and $H_2O$ were used as metal precursor and oxidant respectively. The SIS cycle consisted in subsequent precursor pulses each followed by an exposure step during which the ALD chamber was isolated from the pumping line and the sample was immersed in the precursor vapor. Purging intervals under 100 sccm $N_2$ flow between TMA and $H_2O$ pulse/exposure steps were performed. We started investigating a 10-SIS-cycle process as that we used for infiltrating PS-*b*-PMMA BCP with perpendicular lamellae morphology reported in Ref. [22], then we focused on the first infiltration cycle in pristine PMMA. In the 10-SIS-cycle process, each cycle was made of 0.025 s TMA pulse/60 s exposure/60 s purge, followed by 0.015 s $H_2O$ pulse/60 s exposure/180 s purge. The single cycle in pristine PMMA was investigated as a function of TMA pulse, TMA exposure length, and TMA purge step. After SIS process, the samples were ashed in $O_2$ plasma (40 W, 525 Torr for 10 min), that removed the polymer matrix, leaving alumina films on the Si substrate.

The SIS process was monitored via a rotating compensator ellipsometer equipped with a Xe lamp (M-2000F, J. A. Woollam Co. Inc)[33]. Two quartz windows installed on the ALD reactor lid allowed sending the incident light beam onto the sample and then detecting the reflected light beam at 70° fixed angle with respect to the substrate plane normal. Ellipsometric data Ψ and Δ were collected over the wavelength range from 250 to 1000 nm throughout the entire SIS process time, with an acquisition time of 2.5 s. Short acquisition time were required for dynamic measurements as the polymer films changed continuously during the SIS sequence precursor pulse/exposure/purge. In particular, the polymer modifications as the TMA or $H_2O$ is injected in the ALD chamber are very fast, therefore, the acquisition time was reduced to 1.6 s when monitoring the single first infiltration cycle in pristine PMMA. Data were analyzed to determine the film thickness and refractive index using the EASE software package 2.3 version (J.A. Woollam Co. Inc.)[33], by fitting the ellipsometric data using a film stack model composed of a Cauchy layer model for the polymer film on 2-nm-thick $SiO_2$ on the silicon substrate[34]. *Ex situ* experimental Ψ and Δ spectra of final alumina layers, after $O_2$ plasma, were collected by the same ellipsometer at fixed 75° incidence angle and modeled using the Bruggeman effective medium approximation (EMA), to take into account film porosity and roughness.

The structural morphology of infiltrated polymer films after the $O_2$ plasma step was characterized by field emission scanning electron microscopy (FE-SEM, SUPRA 40, Zeiss) using an in-lens detector and an acceleration voltage of 15 kV.

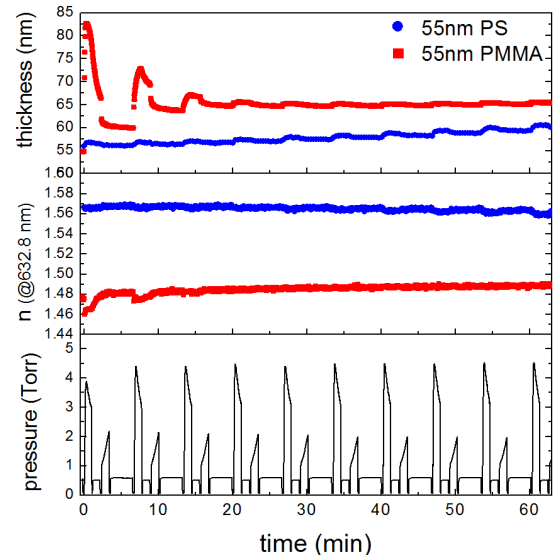

**Figure 1.** Thickness and refractive index n (at 632.8 nm) variation of a PMMA film (square dots) and a PS film (circle dots) during 10-SIS-cycle process. In the bottom panel, pressure vs time showing TMA exposure/purge/ $H_2O$ exposure/purge sequence in each cycle.

## III. RESULTS AND DISCUSSION
### Infiltration in PMMA and PS films

Figure 1 shows the temporal evolution of thickness and refractive index, as extracted from dynamic SE data, for 55-nm-thick PMMA and PS films during 10-SIS-cycle process. The two polymer films exhibit completely different swelling behavior during infiltration. In the first cycle, PMMA thickness largely increases during TMA injection and exposure and subsequently decreases during the following TMA purge and $H_2O$ pulse. Correspondingly, the refractive index decreases during TMA infiltration and then increases back in the remaining part of the SIS cycle. In the second and third cycles, thickness changes are still quite large during TMA injection and exposure steps but less than for the pristine PMMA. Starting from the fourth cycle on, a very small thickness increase is observed during each TMA exposure step. Conversely, PS film exhibits no significant swelling and refractive index changes during the whole infiltration process. The difference in the swelling behavior of PMMA and PS is a signature of the different solubility of TMA in the two polymers. The solubility of TMA in polymers containing C=O groups, as PMMA, is larger than the solubility of TMA in polymers without C=O groups, such as PS, because of the effective intermolecular interaction between TMA and C=O groups, commonly identified in a Lewis acid−base interaction[35].

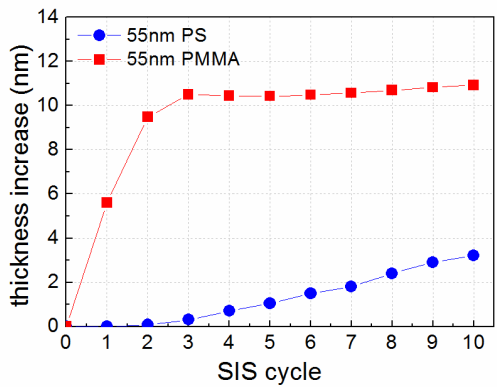

**Figure 2**. PMMA and PS film thickness change during 10-SIS-cycle process.

Starting from collected SE data, we argue that the initial PMMA thickness increase during first TMA exposure is due to TMA molecules diffusing in the polymer film and in small percentage forming an adduct with the carbonyl groups of PMMA. Such physisorbed state is reported to be mainly reversible[18], consequently physisorbed molecules may diffuse out, together with non-reacted TMA molecules during sufficiently long TMA purging steps, resulting in polymer deswelling. During $H_2O$ pulsing, the physisorbed TMA is converted in alumina, fixing $Al_2O_3$ seeds inside the polymer matrix and resulting in an increase of the polymer film thickness at the end of the first SIS cycle. In the following two cycles, TMA diffuses in a polymer matrix including still reachable and not yet saturated C=O moieties together with alumina nucleation sites created in the previous cycle. Consequently, the observed film swelling is lower than in the first cycle, with a corresponding lower thickness increase after $H_2O$ pulse. As the SIS process proceeds, PMMA thickness increase after each cycle is further reduced, suggesting a different growth mechanism associated to the reduced availability of C=O reactive sites and lowered TMA diffusivity inside the polymer.

Interestingly, a small reduction (1.3%) of the refractive index of the infiltrated PMMA film is observed during the large film swelling (50%) occurring throughout the first TMA exposure.

Moreover, almost no variation of the refractive index is registered upon a 20% increment of the PMMA film thickness at the end of the 10-SIS-cycle process. These experimental results indicate that the infiltration of a large amount of aluminum-containing precursor, subsequently converted in alumina with higher refractive index than that of PMMA[36], balances the density reduction expected as a consequence of the large volume increase. As previously discussed, the different swelling behavior of PS films during TMA infiltration, with no significant change in the refractive index, can be ascribed to the absence of reactive carbonyl moieties in PS, with the Al precursor diffusing in and out of the polymer film during each cycle. Nevertheless, a slight but gradual increment in thickness is observed over repeated SIS cycles.

Figure 2 summarizes the PMMA and PS thickness increases at the end of each cycle during the 10-SIS-cycle processes. Despite the inert nature of the polymer matrix with respect to TMA, the PS film shows an appreciable thickness evolution. The growth rate is very similar to that of ALD alumina at 90°C during the first SIS cycles. During the following cycles, the growth rate slightly increases suggesting the occurrence of some oxide growth inside the volume of the PS matrix. The PMMA growth trend is dramatically different, showing two distinct regimes with a large thickness increase during the initial cycles followed by an almost negligible thickness increase in the subsequent cycles of the process.

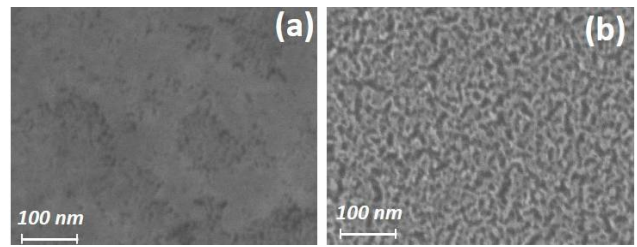

**Figure 3**. FE-SEM images of alumina films resulting from 10-SIS-cycle TMA/$H_2O$ SIS process in 55-nm-thick PMMA (a) and 55-nm-thick PS (b) films after removal of polymer matrix in $O_2$ plasma.

After selective removal of the infiltrated polymer matrix by $O_2$ plasma treatment, the resulting alumina film obtained from PMMA infiltration is 26.9-nm-thick and exhibits refractive index n = 1.47 at 632.8 nm with a smooth and homogeneous surface (Figure 3a). It is worth to note that the refractive index is lower than that of alumina grown by ALD at T<100°C. Actually, a refractive index n = 1.54 was obtained for $Al_2O_3$ films grown by a conventional ALD process at 90°C, in agreement with values reported in literature[37,38]. The low value of the refractive index corresponds to a lower density of the $Al_2O_3$ film and indicates that the film is slightly porous, with a porosity of 16% as evaluated by using the EMA approximation for modelling the optical constants of the film.

From infiltration of the 55-nm-thick PS film, a much thinner (11.4 nm) metal oxide film was obtained compared to that from infiltrated PMMA. This $Al_2O_3$ film exhibits even lower refractive index (n = 1.32) corresponding to a larger porosity of 35%, and a completely different film morphology as shown in Figure 3b. Interestingly, very similar PS thickness increase over 10-SIS-cycle process and similar metal oxide films after polymer removal in $O_2$ plasma were obtained, increasing the duration of the purge step from 60 up to 300 s (Supporting Information S1).

Data indicate that the observed $Al_2O_3$ growth in the PS matrix can not be explained in terms of residual TMA precursor molecules still present in the polymer matrix after a short 60 s purging time. The negligile variation in the final thickness and morphology of the $Al_2O_3$ films observed by increasing the duration of the purge step suggests that the growth of alumina inside the PS film is related to the presence of some defects in the films acting as reaction sites for TMA molecules. These results are of large interest for application of SIS in BCP films, when selective infiltration of PMMA domains respect to PS domains are required in order to replicate the BCP templates. Actually, they indicate that alumina grows inside the PS phase of the BCP matrix, if many SIS cycles are performed and defects are present as nucleation for the growth, independently of the purging length.

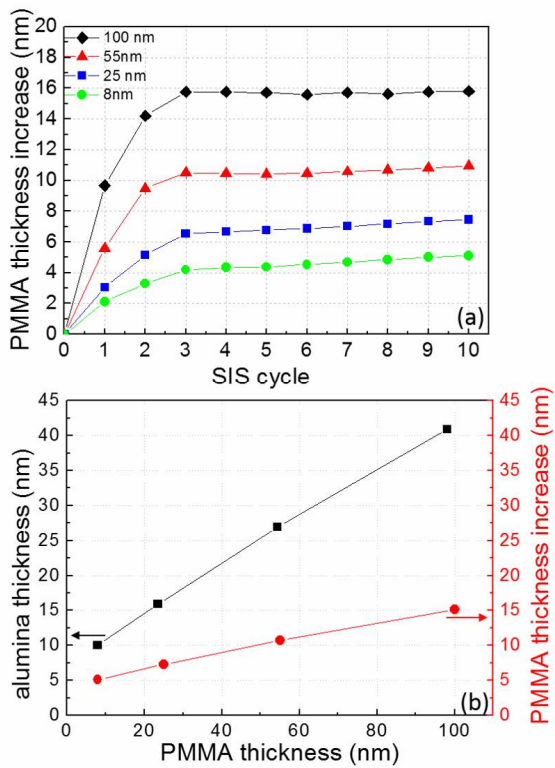

**Figure 4.** (a) PMMA ($M_n$ = 14 kg·mol$^{-1}$) film thickness change at the end of each cycle during 10-SIS-cycle process as a function of initial PMMA film thickness; (b) alumina thickness (square dots) and PMMA thickness increase (circle dots) resulting from the 10-SIS-cycle process as a function of initial PMMA film thickness.

### Infiltration in PMMA films with different thicknesses

Figure 4a reports the thickness increase of PMMA films ($M_n$ = 14 kg·mol$^{-1}$) with initial thickness 100, 55, 25, and 8 nm after each SIS cycle. Independently of the PMMA initial thickness, we observed the occurring of two growth regimes for all PMMA films. During the first 3 SIS cycles, large polymer thickness increase is attributed to large incorporation of metal oxide in the polymer bulk. In the second part of the process the reduced polymer thickness increase may be due to a superposition of several processes: the change in the mechanical properties of the polymer matrix once it has been infiltrated and inorganic material has been grown inside it[4,39], the reduced interaction between TMA and the polymer, being the functional moieties in PMMA largely consumed during the first infiltration cycles, or an effectively reduced incorporation of alumina, assuming TMA diffusion in the PMMA volume is limited by previously grown alumina and corresponds to a very small subsurface infiltration or to an ALD-like growth on the surface in the second part of the infiltration process. Such trend is in agreement with data about alumina infiltrated in random copolymer films as a function of SIS cycle reported in the literature[22], where the largest alumina infiltration is the result of the first SIS cycles, with a half rate reduction in the following cycles.

In Figure 4b, we plot the PMMA thickness increase (right axis) as measured by *in situ* SE at end of the 10-SIS-cycle process as a function of initial PMMA film thickness. We observe that the final infiltrated PMMA thickness increment grows linearly with the starting PMMA thickness. Furtherly, considering the thickness of alumina obtained for the full 10-SIS-cycle process as a function of the initial PMMA thickness we observe that the alumina thickness increases almost linearly but with different rate respect to PMMA thickness increment. The incorporated amount of alumina as a percentage of PMMA initial thickness decreases. We can argue that the film results only partially infiltrated as PMMA thickness increases. In any case, the alumina to PMMA initial thickness ratio remains larger than that reported in Ref. [12], where the calculated ratio for reaction-limited growth inside the polymer is 0.22. We assume that the oxide incorporated in the polymer film during the 10-SIS-cycle process is due to both TMA molecules physisorbed to functional PMMA groups and non-reacted TMA molecules still present inside the polymer after purging and reacting with $H_2O$. In the next paragraph, we analyze the swelling of pristine PMMA films as a function of TMA dose and purging time during the first SIS cycle, and we demonstrate that the proposed picture of the infiltration process is fully supported by *in situ* dynamic SE analysis.

### Evolution of PMMA film during the first SIS cycle

*In situ* SE analysis of SIS process enables real time monitoring of thickness evolution of the PMMA films, providing clear evidence of polymer film modification occurring during the first SIS cycle as a consequence of film exposure to TMA and $H_2O$ molecules. The complex phenomena occurring during the first SIS cycle are extremely interesting, and their comprehension is absolutely necessary because nucleation of $Al_2O_3$ seeds is often used for the subsequent infiltration of precursors that do not present good solubility in PMMA[15]. In particular, TMA exposure phase is characterized by two main process parameters, the TMA dosing, depending in our reactor by the TMA pulse length, and the exposure duration, that can be tuned to allow complete diffusion of the penetrant molecules inside very thick polymer films. Effects of these process parameters on the PMMA film thickness evolution have to be systematically monitored to get information on TMA precursor diffusion and reaction with the host matrix.

For a fixed TMA pulse length and TMA exposure duration, we investigated TMA infiltration in PMMA films with thicknesses ranging from 4 to 100 nm, and with two different molecular weights, corresponding to $M_n$ = 14 kg·mol$^{-1}$ and $M_n$ = 3.9 kg·mol$^{-1}$ respectively. Figure 5 reports the PMMA film thickness evolution recorded during the TMA exposure phase for the various PMMA films with different initial thicknesses. Interestingly, regardless of the initial film thickness or the PMMA molecular weight value, all the samples exhibit the same general behavior, undergoing a significant thickness increase when exposed to TMA and reaching a saturation level.

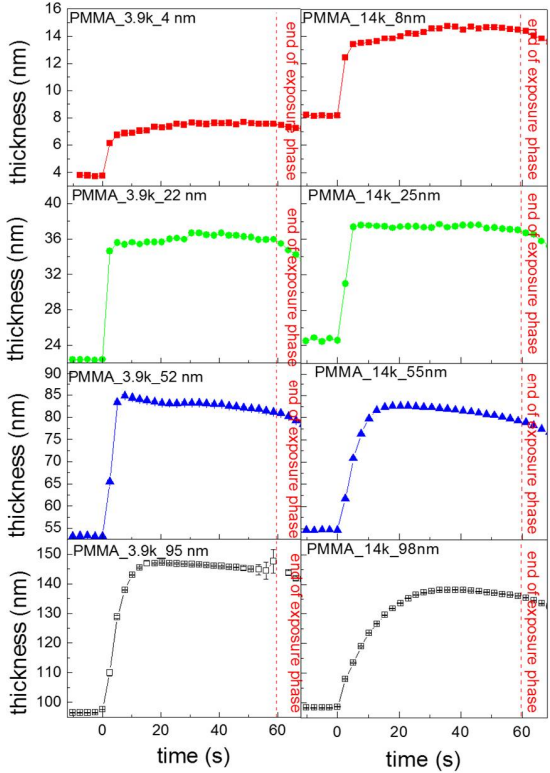

**Figure 5.** Thickness increase of PMMA films during 60 s TMA exposure step in the first SIS cycle for PMMA initial thickness ranging from 4 to 100 nm and PMMA molecular weight $M_n = 3.9$ kg·mol$^{-1}$ and $M_n = 14$ kg·mol$^{-1}$.

Such saturation level is reached faster for thinner samples, since the TMA molecules diffuse over a considerably shorter path. Indeed, it is well known that the timescale of a diffusion process in a layer is extremely influenced by the layer thickness itself, with a dependency $\propto t^{1/2}$ [40]. Moreover, a further increase in thickness is observed in the second part of TMA exposure phase for thinner films, as a consequence of the initial fast sorption of TMA molecules and polymer matrix relaxation, as described below. It is also worth to note that the thickness increase of the thinnest films can be affected by interface effects due to interaction of TMA molecules with the substrate, that can lead to even higher swelling ratio[41]. Finally, the saturation level is reached more quickly in the PMMA films with lower molecular weight, indicating that TMA molecules benefit from a higher diffusivity in polymers with lower molecular weight.

The time evolution of polymeric film swelling $\varepsilon(t)$, defined as

$$\varepsilon(t) = \frac{l(t) - L}{L} \quad (1)$$

where $L$ and $l(t)$ are the initial thickness and the thickness of the polymer film at time $t$ respectively, provides information about the kinetics of the diffusive process of the infiltrating molecules inside the polymeric matrix[42]. For instance, by measuring the dilation of a glassy polyimide thin film caused by the sorption of $CO_2$ molecules at low pressure, Wessling et al.[42] observed a swelling curve following the exactly same evolution of the sorption curves obtained by mass-uptake measurements.

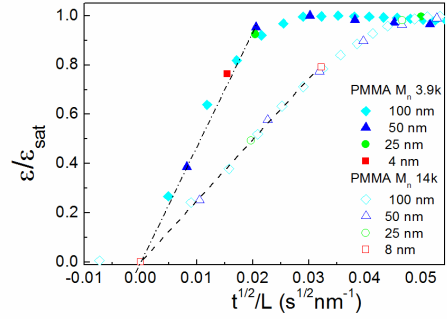

**Figure 6.** $\varepsilon(t)/\varepsilon_{Sat}$ curves as a function of $t^{1/2}/L$ for all the analyzed PMMA samples for TMA exposure step of first SIS cycle.

The initial thickness increase shows a time dependency $\propto t^{1/2}$ and is attributed to the diffusion of penetrant molecules following a Fickian behavior. This first sorption of molecules can produce a stress into the polymeric matrix leading to a slow relaxation of the glassy film. As a consequence of this structural relaxation of the polymeric matrix, more molecules can diffuse and be sorbed into the film, causing a slower secondary increase of mass and thickness of the polymer film[43].

Following this approach, the thickness increase of PMMA films in Figure 5 was analyzed, having in mind that the infiltration process of TMA in PMMA is affected by the fact that TMA molecules fast react to functional groups in the polymer. Figure 6 reports the normalized swelling $\varepsilon(t)/\varepsilon_{Sat}$ as a function of $t^{1/2}/L$ for the various PMMA films. The value $\varepsilon_{Sat}$ corresponds to the maximum swelling reached during the precursor exposure. The linear dependence of $\varepsilon(t)/\varepsilon_{Sat}$ versus $t^{1/2}/L$ during the initial stages of the swelling process is in perfect agreement with equation (2), that approximates the evolution of the swelling $\varepsilon(t)$ for small $t$ (Supporting Information S2), when the infiltration process is dominated by a Fickian diffusion behavior for the precursor molecules infiltrating the polymer [40]:

$$\varepsilon(t) \sim 4\varepsilon_{Sat}\sqrt{\frac{D}{\pi}}\sqrt{\frac{t}{L^2}} \quad (2)$$

where $D$ is the "effective" diffusion coefficient of the precursor molecules in the specific polymer matrix, being precursor diffusion reduced by the reaction of a portion of the precursor molecules when entering the polymer film[5]. From the trends of $\varepsilon(t)/\varepsilon_{Sat}$ with respect to $t^{1/2}/L$ in Figure 6, it is clearly possible to distinguish two separate groups of curves corresponding to the different PMMA molecular weights. They both follow a linear trend for $\varepsilon(t)/\varepsilon_{Sat}$ values up to 0.7, proving that the infiltration is governed by a Fickian diffusion stage in the beginning of the TMA exposure phase. By considering the Fickian part of the curves, the effective diffusion coefficient of TMA molecules inside PMMA samples were calculated using equation (2) through a linear fitting of the experimental data. The corresponding effective diffusion coefficient are $D_{3.9} = 4.5 \times 10^{-12}$ cm$^2$·s$^{-1}$ and $D_{14} = 1.2 \times 10^{-12}$ cm$^2$·s$^{-1}$ for PMMA films with $M_n = 3.9$ kg·mol$^{-1}$ and $M_n = 14$ kg·mol$^{-1}$ respectively. These values confirm that TMA molecules diffuse faster in PMMA samples with lower molecular weight. Interestingly, assuming TMA molecule mean diameter $d \sim 0.5$ nm, the extracted $D$ values are in excellent agreement with diffusivity data available in the literature[44], showing that diffusion coefficients of organic vapors

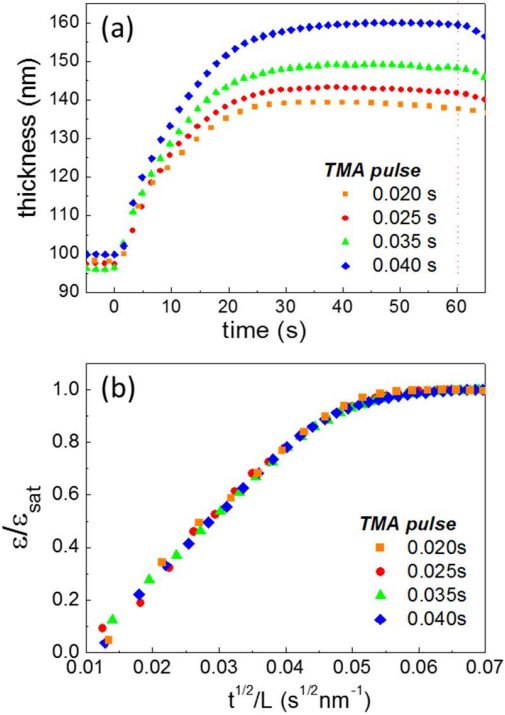

**Figure 7.** (a) Thickness increase of 100-nm-thick PMMA ($M_n$ = 14 kg·mol$^{-1}$) films during 60 s TMA exposure step in the first SIS cycle for increasing TMA dosing; (b) $\varepsilon(t)/\varepsilon_{Sat}$ curves as a function of $t^{1/2}/L$ extracted from curves in (a).

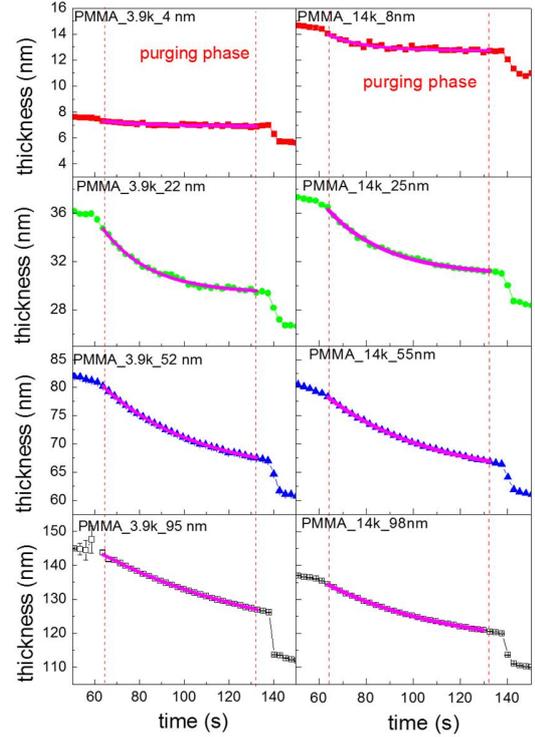

**Figure 8.** Thickness decrease of PMMA films during TMA purging step in the first SIS cycle for PMMA initial thickness ranging from 4 to 100 nm and PMMA molecular weight $M_n$ = 3.9 kg·mol$^{-1}$ and $M_n$ = 14 kg·mol$^{-1}$. (The purging step, corresponding to 100 sccm $N_2$ flow across the reactor chamber, is 60 s long. It is preceded by an interval during which the pumping valve is opened and the chamber evacuated from TMA, and followed by an interval in which the $N_2$ flow is set to zero, the chamber is evacuated again before closing the pumping valve and injecting $H_2O$).

and gases in PMMA at 90°C scale as a function of the penetrant molecule mean diameter $d$.

Similarly, the effect of TMA dosing on the thickness increase of 100-nm-thick PMMA films was monitored by *in situ* dynamic SE (Figure 7a). The thickness increase as a consequence of TMA injection and exposure depends on TMA dosing, varying linearly from 40% to a maximum of 60% of the initial thickness when increasing TMA pulse from 0.02 to 0.04 s. Data indicate that TMA concentration within the polymer matrix depends on the TMA pressure in the reactor and can be tuned by properly adjusting the duration of the TMA pulse. Figure 7b reports the normalized swelling $\varepsilon(t)/\varepsilon_{Sat}$ as a function of $t^{1/2}/L$ for increasing TMA dosage. All the curves exhibit the same linear behavior during the initial stages of the process, irrespective of the TMA dosing. Experimental data indicate that the TMA infiltration is characterized by an initial Fickian diffusion regime with almost the same effective diffusion constant that is independent on the TMA dosing in the range of values explored in this work.

Continuing the monitoring of the thickness evolution of the PMMA films during the purging step in the first SIS cycle after TMA exposure, a significant decrease of PMMA film thickness is observed (Figure 8), as the reactor is vacuum pumped and then crossed by a 100-sccm $N_2$ flow for 60 s. Thickness evolution as a function of time follows an exponential trend $e^{-t/\tau}$ where $\tau$ is assumed to be the effective time constant of the desorption process. Figure 9 reports $\tau$ values extracted from curves in Figure 8, as function of the PMMA film thickness. A progressive increase of $\tau$ values is observed when increasing PMMA film thickness. Interestingly, similar $\tau$ values were measured for polymer films with approximately the same thickness, independently of the PMMA molecular weight. During the purging phase, out-diffusion of non-reacted TMA molecules and desorption and loss of a fraction of TMA molecules that are reversibly physisorbed to C=O groups are expected to occur, producing the shrinking of the polymer matrix. Moreover, the polymer film is subjected to internal stresses that can further contribute to the observed decrease of the film thickness. Therefore, competing processes with different characteristic times occur simultaneously during the desorption process.

The $\tau$ values extracted from SE monitoring of the deswelling process (Figure 9) are much shorter than those evaluated from FTIR analysis. Actually, $\tau$ values of the order of tens of minutes are obtained monitoring C=O absorbance peak intensity variation upon increasing purging time, as reported by Biswas et al.[17,18]. Biswas *et al* assigned the temporal evolution of C=O absorption intensity to desorption of physisorbed TMA from the reversible adducts C=O···Al(CH$_3$)$_3$ formed during TMA exposure phase. It is worth noting that the time scale of FTIR investigation is different respect to that of SE analysis. By *in situ* dynamic SE the polymer deswelling process is monitored during removal of TMA vapor from the reactor and purging in $N_2$ flow. Conversely, even in the case of the shortest purge step of 35 s, the first FTIR spectrum was obtained as the average over a 320-s-long measurement[17,18]. Therefore, we argue that the observed thickness evolution is mainly associated to the out-diffusion of non-reacted TMA molecules from the PMMA films.

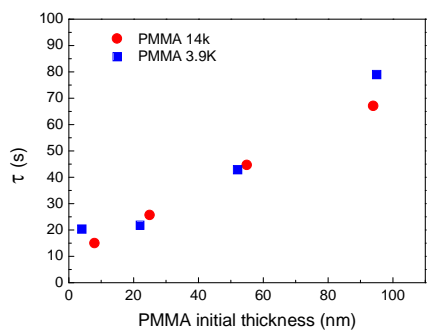

**Figure 9.** Characteristic time of PMMA deswelling during purging step in the first SIS cycle as a function of initial film thickness.

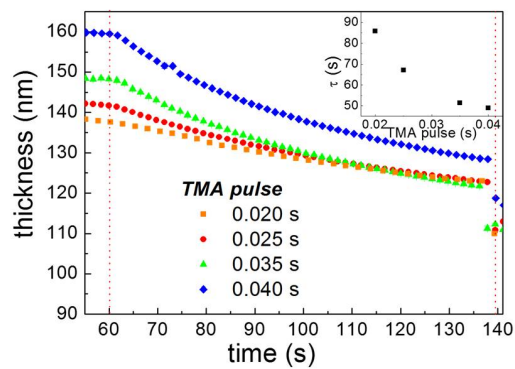

**Figure 10.** Thickness change of 100-nm-thick PMMA ($M_n$ = 14 kg·mol$^{-1}$) films during TMA purge step in the first SIS cycle for increasing TMA dosing. In the inset the correspondent characteristic time $\tau$.

According to the experimental data, the out-diffusion process is significantly slower than that of the infiltration of TMA molecules in pristine PMMA matrix. Moreover, the out-diffusion time constants are clearly dependent on the thickness of the polymer film and almost independent of the molecular weight, differently from what observed during TMA exposure and infiltration. This behavior can be attributed to the fact the infiltrated polymeric matrix is significantly different compared to the pristine PMMA matrix. Considering the process occurring in unperturbed PMMA matrix, structural relaxation characteristic times are expected to be independent of the polymer film thickness, since they are expected to be strictly related to the elastic properties of the polymer matrix. On the contrary, shorter $\tau$ values for deswelling were evaluated for thinner PMMA films during purging phase. We could argue that the polymer matrix is modified by the TMA infiltration, in larger extent for thinner films, and that the observed deswelling depends on the specific history of polymer systems during the infiltration process.

Figure 10 depicts the thickness evolution during the purging step in the case of 100-nm-thick PMMA films infiltrated increasing input TMA doses by increasing the TMA pulse duration. From fitting of the experimental data, the time constant of the deswelling process were obtained. The inset of Figure 10 reports the $\tau$ values as a function of the duration of the TMA pulse. Interestingly, the time constant of the exponential thickness decrease is inversely proportional to the TMA dose, i.e. the deswelling process is faster for larger TMA doses, that is for larger sorption-induced polymer dilation.

To get more information about the residual TMA molecules present in the polymer at the end of the purging step, the PMMA shrinking occurring upon H$_2$O injection in the chamber was investigated in details. The reaction of TMA molecules with H$_2$O inside the polymer is extremely fast with an almost instantaneous decrease of the PMMA film thickness upon H$_2$O pulse. This results is perfectly consistent with data on H$_2$O molecules diffusion in a PMMA matrix[44]. The PMMA thickness shrinking upon H$_2$O infiltration and reaction was observed to be in relation with the metal oxide incorporated in the polymer film. For instance, the 100-nm-thick PMMA films infiltrated with different TMA pulses showed very similar shrinking as a consequence of H$_2$O pulse.

After O$_2$ plasma treatment to remove the polymer matrix, resulting alumina films have similar thicknesses, ranging from 14.2 to 17.8 nm for TMA pulse length from 0.02 up to 0.04 s, and very similar refractive index (n = 1.47 at 632.8 nm). Therefore, the larger polymer swelling as a consequence of higher TMA partial pressure in the reactor chamber has a minimal impact on the final amount of incorporated alumina, because of very fast out-diffusion of non-reacted TMA during N$_2$ purging, whereas similar shrinking after reaction with H$_2$O corresponds to similar infiltrated alumina.

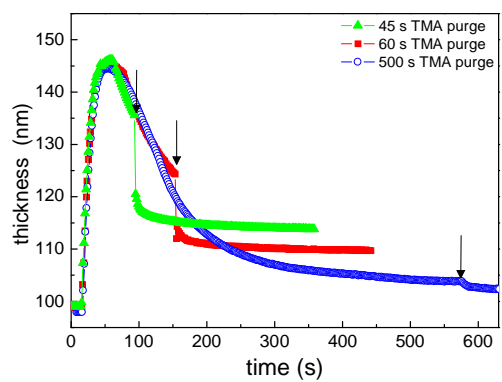

**Figure 11.** Thickness evolution of 100-nm-PMMA films for 1-SIS-cycle with 0.025 s TMA pulse with increasing purge length. The arrows indicate the H$_2$O pulse.

Finally, the effect of the purging time on the final amount of TMA trapped in the PMMA film was investigated by increasing the duration of the purge step from 60 to 500 s. Figure 11 reports the time evolution of the PMMA film thickness during the first SIS cycle for 100-nm-thick PMMA films. The TMA pulse is 0.025 s. When applying a very long N$_2$ purge (500 s), the PMMA film deswells back to a thickness value that represents a sort of limiting value. The final thickness of the PMMA film upon a 500-s-long purge step is about 8% larger than the initial thickness and, after infiltration with H$_2$O molecules, it further reduces to about 5%. We can consider this small polymer thickness increase as associated to the incorporation in the polymer film of alumina grown from TMA molecules chemically reacted with PMMA functional groups, since almost all physisorbed TMA molecules were desorbed away during this very long purge step. FTIR studies demonstrated that the amount of chemisorbed TMA molecules corresponds to a small fraction of the infiltrated and physisorbed TMA molecules[17,18].

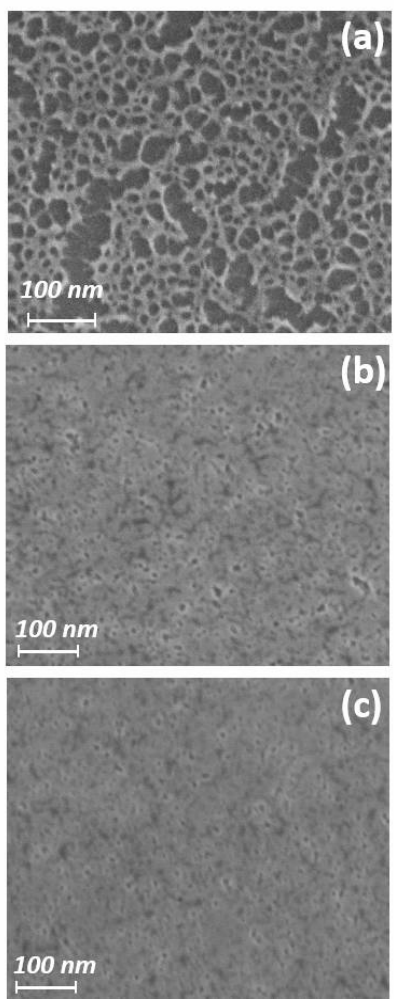

**Figure 12.** FE-SEM micrographs of Al$_2$O$_3$ films from 1-SIS-cycle process with purging time of 500 (a), 60 (b) and 45 s (c) in 100-nm-thick PMMA after removal of polymer matrix in O$_2$ plasma.

Looking at the morphology of the film after removal or the polymer matrix (Figure 12a) we observed that the film is not continuous and with a high level of porosity. Reducing the purge length after TMA exposure to 60 s, the deswelling curve shows that TMA desorption is abruptly interrupted by injection of H$_2$O that reacts to form alumina resulting in a continuous 14.4-nm-thick but still porous film (Figure 12b). The film morphology and the time scale of our observation suggests that we are looking at a film resulting from infiltrated alumina grown by the reaction between a fraction of physisorbed TMA molecules on functional groups in PMMA and H$_2$O. Data clearly indicate that the thickness of the Al$_2$O$_3$ film can be modulated by properly adjusting the process parameters in order to control the amount of TMA molecules effectively inside the polymer film. Then we further reduced the purging time (to 45 s) in order to incorporate more alumina in the polymer film. The correspondent curve in Figure 11 shows an effective larger swelling at the end of both the purging step when H$_2$O is injected in the reactor, and the completed SIS cycle, resulting in an alumina film with similar morphology (Figure 12c) and with same refractive index, but significantly thicker (23.4 nm) than the one obtained with 60-s-purge (14.4 nm). Therefore, the largest swelling ratio between the end of the purging step and the end of SIS cycle, in addition to the final thickness, is an evidence of the largest alumina incorporated in the polymer film.

## IV. CONCLUSIONS

PMMA and PS thin films with different thicknesses have been infiltrated using a SIS process based on TMA and H$_2$O precursors. The infiltration process was monitored by *in situ* dynamic SE and investigated in relation to the resulting alumina films after removal of the polymer matrix by O$_2$ plasma. Different swelling behavior of the two polymers was observed evidencing the larger solubility of TMA in PMMA respect to PS. The resulting Al$_2$O$_3$ films was thicker and denser in the case of infiltrated PMMA with respect to PS film. PMMA thickness increase after 10-SIS-cycle process, as a function of initial polymer film thickness, was related to the final alumina properties. Intra-cycle investigations of swelling and deswelling during TMA exposure and subsequent purging allowed obtaining useful information on the SIS process. TMA infiltration in PMMA follows a Fickian diffusion behavior with diffusion constant dependent on the PMMA molecular weight. TMA desorption similarly follows the same diffusion behavior but with a slower rate respect to in-diffusion in pristine polymer, as the polymer has been modified by TMA interaction. Changing the purging length after TMA exposure resulted in different swelling of the PMMA films and yielded in alumina films with different morphology and porosity. In this respect, *in situ* dynamic SE offers the possibility to gain time-resolved insight in the infiltration process both in terms of the inorganic material amount retained in the polymer film and of diffusion and sorption or entrapping processes of precursor inside the polymer.

## ASSOCIATED CONTENT

**Supporting Information** (1) PS thickness change as extracted by *in situ* SE monitoring over 10-SIS-cycle process, together with Al$_2$O$_3$ growth on SiO$_2$/Si at 90°C. FE-SEM image of alumina film obtained from 10-SIS-cycle process in 43-nm-thick PS, for increasing TMA purge step, after removal of the polymer in O$_2$ plasma. (2) Mathematical model for the concentration profile of penetrants inside a layer of thickness L - as a function of exposure time – for a Fickian diffusion with a fixed diffusion coefficient D.
This material is available free of charge via the Internet at http://pubs.acs.org.

## AUTHOR INFORMATION

**Corresponding Author**
*(E.C.) E-mail: elena.cianci@mdm.imm.cnr.it
**Author Contributions**
The manuscript was written through contributions of all authors.
**Notes**
The authors declare no competing financial interest. Patent protection related to this work is pending (International Patent Application No. PCT/IB2014/061324).

## ACKNOWLEDGMENTS
The authors would like to acknowledge Jacopo Frascaroli, Tommaso Giammaria, Fabio Zanenga (CNR, Italy) and Michele Laus (UniPO, Italy) for fruitful discussions, Katia Sparnacci (UniPO, Italy) for the synthesis of the PS and PMMA polymers and Mario Alia (CNR, Italy) for technical assistance. This research has been partially supported by the European Union's Horizon 2020 research and innovation program under grant agreement No 688072 "IONS4SET".

For table of contents (TOC) only

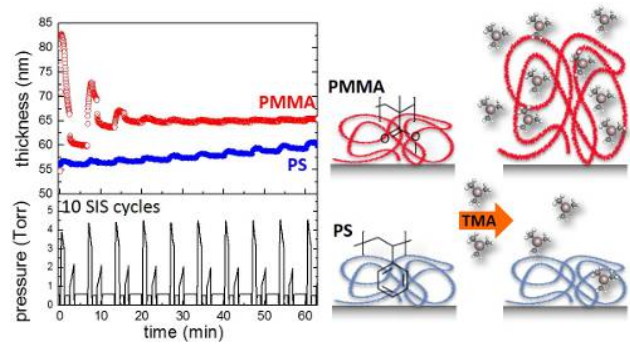